\documentclass[aps,pra,twocolumn,10pt]{revtex4-1}
\usepackage{graphicx}
\usepackage{amsmath}
\usepackage{amssymb}
\usepackage{mathrsfs}
\usepackage{amsfonts}
\usepackage{dsfont}
\usepackage{dcolumn}% Align table columns on decimal point
\usepackage{bm}% bold math
\usepackage{booktabs}
\usepackage{tabularx}
\usepackage{textcomp}
\usepackage{color}
\usepackage{xcolor}

\everymath{\displaystyle}
\providecommand{\abs}[1]{\left\lvert#1\right\rvert}

\newcommand{\br}{\mathbf{r}}
\newcommand{\bx}{\mathbf{x}}
\newcommand{\by}{\mathbf{y}}
\newcommand{\bz}{\mathbf{z}}
\newcommand{\bp}{\mathbf{p}}
\newcommand{\bq}{\mathbf{q}}
\newcommand{\bn}{\mathbf{n}}

\newcommand{\bk}{\mathbf{k}}
\newcommand{\vA}{\mathbf{A}}
\newcommand{\vE}{\mathbf{E}}
\newcommand{\vB}{\mathbf{B}}

\newcommand{\vP}{\mathbf{P}}
\newcommand{\vS}{\mathbf{S}}
\newcommand{\bnabla}{\bm{\nabla}}

\usepackage[all]{xy}
\begin{document}
%
%\title{Generating light with transverse spin}
\title{Transverse spin of light for all wavefields}
\author{Andrea Aiello$^{1,2}$}
\email{andrea.aiello@mpl.mpg.de}
\author{Peter Banzer$^{1,2}$}
%\author{Gerd Leuchs$^{1,2,3}$}
%
\affiliation{$^1$Max Planck Institute for the Science of Light, G$\ddot{u}$nther-Scharowsky-Strasse 1/Bau 24, 91058 Erlangen, Germany}
\affiliation{$^2$Institute for Optics, Information and Photonics, University of Erlangen-Nuernberg, Staudtstrasse 7/B2, 91058 Erlangen, Germany}
%\affiliation{$^3$ Department of Physics, University of Ottawa, 25 Templeton, Ottawa, Ontario K1N 6N5 Canada}
%
\date{\today}
\begin{abstract}
It has been known for a  long time that light carries both linear and angular momenta parallel to the direction of propagation. However, only recently it has been pointed out that beams of light, under certain conditions, may exhibit a transverse spin angular momentum perpendicular to the propagation direction. When this happens, the electric field transported by the light rotates around an axis transverse to the beam path. Such kind of fields, although deceptively elusive,  are almost ubiquitous in optics as they manifests  in strongly focused beams, plasmonic fields  and evanescent waves. In this work we present a general formalism describing all these phenomena. In particular, we demonstrate how to mathematically generate a wave field possessing transverse spin angular momentum density, from any arbitrarily given scalar wave field, either propagating and evanescent.
\end{abstract}
\maketitle

\section*{Introduction}\label{Intro}

The electric field of a monochromatic electromagnetic wave of angular frequency $\omega$ spans, at any fixed point in space,  an elliptical trajectory in a period $T= 2 \pi/\omega$ \cite{BW}. In the points where the ellipse reduces to a circle, %
%In a plane-wave solutions of Maxwell equations, the electric and magnetic fields are orthogonal each other and both are %perpendicular to the wave vector $\bk$. When the electric field carried by these waves rotates with uniform angular %frequency $\omega = c \abs{\bk}$ around the $\bk$-axis,
the light is said to be circularly polarized. When the light propagates in the direction perpendicular to the plane of such circles, the uniform motion of the electric field vector generates a \emph{longitudinal} spin angular momentum, loosely dubbed ``helicity'' \cite{NoteHelicity}. However, there are instances where the light is circularly polarized in a plane containing the direction of propagation. In these cases the light carries a \emph{transverse} spin angular momentum density \cite{PhysRevLett.103.100401,PhysRevA.81.053838,JEOS:RP13032}. When the light wave exhibits both  transverse and longitudinal electric field components, the transverse spin angular momentum can be either dependent or independent of the longitudinal one \cite{PhysRevA.85.061801,BliokhEvanescent,Bliokhtwowaves,doi:10.1021/nl5003526}.

%At any given point $\br$, there is a  unique plane perpendicular to the propagation direction of the light. The motion of the electric field vector upon such plane determines the so-called ``transverse polarization'' of the light. Similarly, the electric field vector kinematics upon the only plane through $\br$ and containing the axis of propagation, fixes the ``longitudinal polarization''.

In this work we aim at establishing a perfectly general theory of wave fields displaying transverse spin angular momentum (AM). We begin with studying  particular solutions of  Maxwell's equations possessing both transverse and longitudinal components of the electric field vector. The kinematics of the latter is investigated by means of suitably defined Stokes parameters \cite{Jackson}.
 This parametrization permits us to establish general sufficient conditions for the existence of transverse spin AM.
 Under certain circumstances, such conditions coincide with the  Cauchy-Riemann equations satisfied by holomorphic complex functions \cite{Lang}.  This novel result establishes an intriguing and previously unnoticed connection between the theory of complex functions and the transverse spin AM of light. Finally, to illustrate our findings,
we work out two detailed examples involving non-diffracting Bessel beams and evanescent waves. \emph{Caveat}: In the remainder, we adopt units where the electric field has the dimensions of a wave-number, namely the inverse of a length.

\section*{Transverse spin angular momentum}\label{Theory}

In the Coulomb gauge \cite{Jackson} the electromagnetic vector potential  $\vA(\br,t) $ is purely transverse, namely $\bnabla \cdot \vA=0$, and satisfies the wave equation $\Box \vA = 0$. A  plane-wave solution of these equations takes the form $\vA(\br,t) = \mathbf{n} \exp(i \bk \cdot \br - i\omega t)$, where $\bk$ is the either real- or complex-valued wave vector \cite{Jackson}, $\omega = c \abs{\bk} \equiv c k$ is the angular frequency and $ \bn$ is a constant three-vectors such that $\bk \cdot \bn =0$. In the Cartesian reference frame $(x,y,z)$ the latter condition can be written as
\begin{align}\label{kdotn}
\bk \cdot \bn = k_x n_x + k_y n_y + k_z n_z = 0.
\end{align}
Apart from the trivial solution $\bn = (0,0,0)$,  Eq. \eqref{kdotn} also admits  three elementary  solutions of the form $\bn_1 = (-k_y,k_x,0),\bn_2 = (-k_z,0,k_x),\bn_3 = (0,-k_z,k_y)$, where one of the three Cartesian components of $\bn$ is chosen to be zero. The first solution $\bn_1$ generates a plane-wave mode whose electric field is purely perpendicular to the $z$-axis \cite{Lekner01}. These modes are commonplace in the theory of electromagnetic waveguides \cite{MIT}. For instance, since the condition $\bk \cdot \bn_1 = 0$ is fulfilled irrespective of the value of $k_z$, one can choose $\bk = \bk_1 = (k_x,k_y,0)$ and obtain the so-called transverse electric (TE) modes of a rectangular waveguide \cite{Smith11}.
The remaining two solutions $\bn_2, \bn_3$ generates plane-wave modes with both longitudinal ($z$) and transverse (either $x$ or $y$, respectively) components of the electric field. As an example,
when choosing $\bk = \bk_2 = (k_x,0,k_z)$ and $\bk = \bk_3 = (0,k_y,k_z)$ with $\bn = \bn_2$ and $\bn = \bn_3$, respectively, one attains the transverse magnetic (TM) modes of a rectangular waveguide.

Consider now $\bn = \bn_3$. In this case the vector potential can be written as
\begin{align}\label{vApw}
\vA(\br,t) = & \; (0,-k_z,k_y) \exp(i \bk \cdot \br -i\omega t) \nonumber \\
= & \; -i(0,-\partial_z,\partial_y) \exp(i \bk \cdot \br - i\omega t).
\end{align}
 Let $\psi(\br,t)$ be an arbitrary  solution of the wave equation, namely $\Box \psi = 0$. By definition, this field can always be expressed as a superposition of plane waves. Therefore, using Eq. \eqref{vApw} one can write $\vA =   \left(0,- \partial_z \psi, \partial_y \psi \right)$, with $\vA$ automatically satisfying both $\bnabla \cdot \vA=0$ and $\Box \vA = 0$. The electric and magnetic fields generated by $\vA$ are
\begin{align}
\vE =  & \; -\dot{\vA} =  \bigl(0, \partial_z \dot{\psi}, -\partial_y \dot{\psi} \bigr), \label{E} \\
\vB =  & \; \bnabla \times \vA =  \bigl(\partial^2_y \psi +\partial^2_z \psi, -\partial_x \partial_y \psi, -\partial_x \partial_z \psi  \bigr), \label{B}
\end{align}
where, $\dot{\vA} = \partial_t \vA$ and $\dot{\psi} = \partial_t \psi$.  It is interesting to note that the electromagnetic fields above can also be obtained from the Hertz magnetic vector potential $\bm{\Pi}_m = \left( -\psi/\mu_0,0,0 \right)$, with $\vA = \mu_0\bnabla \times  \bm{\Pi}_m$  \cite{Jackson,Hertz}.

When $\psi$ represents a time-harmonic wave of the form $\psi(\br,t) = U(\br) \exp\left( - i  \omega t \right)/\omega$, the physical electric field vector
\begin{align}\label{RealE}
\vE^R(\br,t) =  & \; \operatorname{Re}\left[ - i  \bigl(0, \partial_z U, -\partial_y U \bigr)\exp\left( - i  \omega t \right) \right] \nonumber \\
=  & \; \mathbf{p}(\br) \cos (\omega t) + \mathbf{q}(\br) \sin (\omega t)
\end{align}
sweeps out an ellipse lying in the $yz$-plane, where $\mathbf{p} = (0, \partial_z v, -\partial_y v)$ and $\mathbf{q} = (0, -\partial_z u, \partial_y u)$, with $u(\br)$ and $v(\br)$ denoting, respectively, the real and the imaginary part of $U=u+iv$ \cite{BW}. When the latter is given,  the magnitude and the orientation of the axes of the ellipse at a fixed point $\br=(x,y,z)$ in space, are determined by the coordinates  $x,y$ and $z$ of the point. In particular, if at $\br=\br_0$ $\bp(\br_0) \cdot \bq(\br_0) = 0$ and $ \bp^2(\br_0) - \bq^2(\br_0) =0$ are fulfilled, then the wave is {circularly polarized} in the $yz$-plane \cite{BW} and the spin AM density is purely transverse at $\br=\br_0$.
In other points on the   $yz$-plane, the polarization  can be completely characterized  by the spatially-varying \emph{Stokes parameters} $S_0(\br),S_1(\br),S_2(\br),S_3(\br)$ defined in terms of the vectors $\bp, \bq$, as
\begin{align}
S_0  =  & \; p_z^2 + q_z^2 + p_y^2 + q_y^2  \nonumber \\
     =  & \; (\partial_y v)^2 + (\partial_y u)^2 + (\partial_z v)^2 + (\partial_z u)^2 ,\label{StokesPQ1} \\
S_1  =  & \; p_z^2 + q_z^2 - p_y^2 - q_y^2 \nonumber \\
     =  & \;  (\partial_y v)^2 + (\partial_y u)^2 - (\partial_z v)^2 - (\partial_z u)^2 , \label{StokesPQ2}  \\
S_2  =  & \; 2 \left( p_z p_y + q_z q_y \right) \nonumber \\
     =  & \; -2 \left( \partial_z u \partial_y u + \partial_z v\partial_y v \right),\label{StokesPQ3} \\
S_3  =  & \; 2 \left( p_z q_y - p_y q_z \right)\nonumber \\
     =  & \; 2  \left( \partial_z u \partial_y v - \partial_y u\partial_z v \right ), \label{StokesPQ4}
\end{align}
where we have used the notation by Dennis \cite{Dennis02}.

 The first constraint $\bp(\br_0) \cdot \bq(\br_0) =0$ for the existence of circular polarization at $\br=\br_0$, can be expressed in terms of $u,v$ as: $\partial_z u \partial_z v + \partial_y u\partial_y v =0 $. A possible solution of this equation is
\begin{align}\label{CR}
\left(\partial_z v \right)(\br_0) = \left(-\partial_y u\right)(\br_0),   \; \; \; \left(\partial_y v\right)(\br_0)  = \left(\partial_z u\right)(\br_0) ,
\end{align}
where the notation emphasizes the fact that the equations in \eqref{CR} are supposed to become equalities only in the isolated point $\br_0$, and \emph{not necessarily everywhere} on the   $yz$-plane. Then at $\br=\br_0$,   $\mathbf{p} = - (0, \partial_y u, \partial_z u)$ and the second condition  $ \bp^2 - \bq^2 =0 $ is \emph{automatically} satisfied. Moreover, from Eqs. (\ref{StokesPQ1}-\ref{StokesPQ4}) it follows that $S_1 = S_2 =0$ and $S_3/S_0 =1$, as expected for a circularly polarized wave.

Equation  \eqref{CR} is the first main result of this work and  establishes a necessary and sufficient condition for the existence of transverse spin AM density in fields of the form (\ref{E}-\ref{B}). This concept can be further extended by taking for
 $U = u+i v$, a \emph{holomorphic function} of the variables $y$ and $z$. Then, the Cauchy-Riemann equations satisfied by $U$ to be holomorphic, read  $\partial_z v = -\partial_y u$ and $\partial_y v = \partial_z u$ \cite{Lang}, which coincides with Eq.  \eqref{CR} at $\br = \br_0$.
Therefore, when $U$ is holomorphic the light carries a transversely spinning electric field (or, synonymously, transverse circular polarization) \emph{uniformly} distributed everywhere on the $yz$-plane.
 Elementary solutions of the monochromatic Helmholtz equation $( \nabla^2 + k^2 )U=0$, which are also holomorphic with respect to the variables $y$ and $z$, must have the form $U(\br) = H(y,z) \exp(i k x)$, where $H$ is an arbitrary harmonic function on the $yz$-plane, namely $(\partial_y^2 + \partial_z^2)H=0$. Of course, by definition of harmonic functions, $H$ cannot be bounded on the $yz$-plane and, therefore, it does not admit a finite norm. However, this does not prevent the existence of finite functions approximating $H$ (see, e.g.,  discussions in \cite{Lekner01,LeknerSheppard}).

 A simple, practical ``recipe'' to mathematically obtain uniform circular polarization everywhere on the $yz$-plane, can be given by simply noting that if $f(\xi)$ is an arbitrary smooth real function with respect to the real variable $\xi$, then $f(z + i y) = a(z,y) + i b(z,y)$ is by definition a holomorphic function with respect to the variables $y$ and $z$ \cite{Lang}. Now, with the harmonic real functions $a$ and $b$, one can build the two reciprocally orthogonal vectors $\bp = (0,a,b)\cos(k x)$ and $\bq = (0,-b,a)\cos(k x)$. Substituting these vectors in Eq. \eqref{RealE}, one obtains an electric field with uniform transverse circular polarization on the $yz$-plane.

The physical content of our theory can be further elucidated by calculating the quantity  $\vE^R \cdot \dot{\vE}^R$, key to the magnetic circular dichroism manifested by molecules in DC magnetic fields \cite{doi:10.1021/jp1092898}. From  Eq. \eqref{RealE}  it follows that  $\vE^R \cdot \dot{\vE}^R = \omega \left( \bp \times \bq\right)$. Comparing this result with Eq. \eqref{StokesPQ4} we immediately find
\begin{align}\label{MCD}
\vE^R \cdot \dot{\vE}^R = -\frac{ \omega}{2} \bigl(S_3,0,0 \bigr).
\end{align}
This means that our wave field \eqref{E} couples to molecules subjected to a DC magnetic field in the $x$ direction and that the coupling is maximum ($S_3 = \pm 1$) at the points where the electric field vector sweeps out a circle on the
$yz$-plane.

The main virtue of the fields of the form (\ref{E}-\ref{B}) resides in their straightforward physical significance that leads to the establishment of an intriguing connection between the distribution of transverse spin AM density in light waves and the theory of complex functions. However, more general wave fields with nonzero longitudinal electric field vector component can be obtained by considering an electromagnetic vector potential of the form $\vA = \vA_2 + \vA_3$, where $\vA_2 = \left(  - \partial_z \phi, 0,\partial_x \phi\right)$ and $\vA_3 = \left( 0, - \partial_z \psi, \partial_y \psi\right)$, with $\phi(\br,t), \psi(\br,t)$ being two independent solutions of the wave equation. This choice yields the electric field vector
\begin{align} \label{Egeneral}
\vE =    \bigl(\partial_z \dot{\phi}, \partial_z \dot{\psi}, -\partial_x \dot{\phi}-\partial_y \dot{\psi} \bigr),
\end{align}
which simultaneously displays both nonzero transverse and longitudinal Cartesian components.  The expression in Eq. \eqref{Egeneral} is interesting because it also implies that a wave of the form \eqref{Egeneral} has necessarily a vanishing longitudinal component when the light is uniformly circularly polarized everywhere in the $xy$-plane. This can be seen again by using complex calculus. Indeed, if we choose $\psi = i \phi$ with  $\phi(\br,t) = U(\br) \exp\left( - i  \omega t \right)/\omega$, then we have $\bp = \left( \partial_z v, \partial_z u, -\partial_y u - \partial_x v\right)$ and $\bq = \left( -\partial_z u, \partial_z v, \partial_x u - \partial_y v\right)$. The conditions $\bp \cdot \bq = 0 = \bp^2 - \bq^2$ to be fulfilled in order to have circular polarization in the plane spanned by $\bp$ and $\bq$, now read $\partial_y v = \partial_x u$, and $\partial_x v = -\partial_y u$. The latter relations coincide with the Cauchy-Riemann equations for holomorphic functions in the variables $x$ and $y$.  When satisfied, these equations lead to $p_z =0 = q_z$, namely to a vanishing longitudinal electric field vector component. This case illustrates once more  the strict connection existing between transverse circular polarization of light and the theory of complex functions.

\section*{Applications of the formalism}\label{Examples}

The  scalar field $\psi(\br,t)$  considered up to now, was completely arbitrary. This choice permitted us to find some perfectly general fundamental relations. However, for the sake of definiteness, we now specialize in the study of a scalar zeroth-order Bessel field and of an evanescent plane wave.

\subsection{Bessel field}\label{Besselfield}

Consider the scalar Bessel field
\begin{align}\label{Bessel}
\psi(\br,t) = J_0\left( r k_{r}  \right) \exp\left(i z  k_z  \right) \frac{\exp \left( - i \omega t \right)}{\omega},
\end{align}
where $k_r = k \sin \vartheta_0$ and $k_z = k \cos \vartheta_0$, with $\vartheta_0$ denoting the aperture of the Bessel cone and $x = r \cos \theta, \;	y = r \sin \theta$ defining the polar coordinates $(r,\theta)$ in the $xy$-plane \cite{PhysRevLett.58.1499}.
From Eq. \eqref{Bessel} it follows that $U =  J_0\left( r k_{r}  \right) \exp\left(i z  k_z  \right)$ with
 $u =  J_0 \left( r k_{r}  \right) \cos\left( z  k_z  \right)$ and $v =  J_0 \left( r k_{r}  \right) \sin \left( z  k_z  \right)$. The conditions for circular polarization Eq. \eqref{CR} reduce therefore to the single equation
\begin{align}\label{CRBessel}
\cot \vartheta_0 = \frac{y}{r} \frac{J_1\left( r k_{r}  \right)}{J_0\left( r k_{r}  \right)} ,
\end{align}
which can be numerically inverted to find the points $\br_0=(x_0,y_0,z)$ where Eq. \eqref{CRBessel} becomes an identity.

The real electric field generated by $\psi(\br,t)$  using Eq. \eqref{E}, can be expressed in polar coordinates $(r,\theta)$ as
\begin{align}\label{ERealBessel}
\vE^R(r,\theta,t) =  & \;  \hat{\by} \, k_z J_0\left( r k_r\right) \cos \left( z k_z - \omega t \right) \nonumber \\
&  + \hat{\bz} \, k_r \sin \theta J_1\left( r k_r\right) \sin \left( z k_z - \omega t \right) .
\end{align}
The temporal evolution of $\vE^R(r,\theta,t)$ as a function of the scaled coordinates $z/z_0$ and $y/r_0$, superimposed on a density plot of the  norm of the field  $(\vE^R \cdot \vE^R )^{1/2}$, is shown in Fig. \ref{fig5} below. In the remainder, $x$ and $y$ coordinates are normalized to $r_0 = a/k_r$, with $a \approx 2.405$ being the smaller solution of the equation $J_0(a)=0$, while the $z$ coordinate is normalized to $z_0 = 1/k_z$. Moreover, in all plots we choose $\vartheta_0 = 15^\circ$.
\begin{figure}[ht!]
\centerline{\includegraphics[scale=3,clip=false,width=1.1\columnwidth,trim = 0 0 0 0]{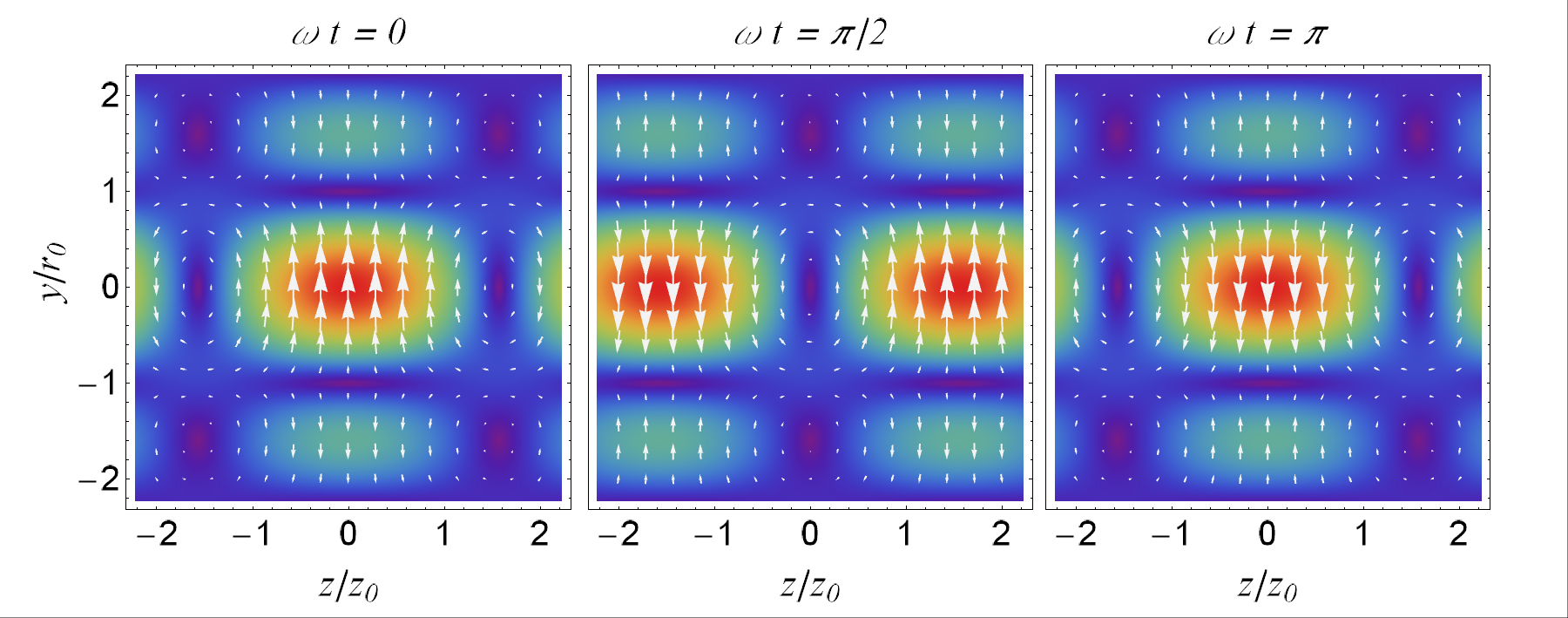}}
 \caption{ \label{fig5}
Instantaneous distributions of the real-valued electric field $\vE^R(r,\theta,t)$  \eqref{ERealBessel}. At
$y / r_0 \approx 1$ the electric field spins in the $yz$-plane. Here  $r_0 = a/k_r$ gives the central core spot size of the beam, with $a \approx 2.405$  such that $J_0(a)=0$ and $\vartheta_0 = 15^\circ$. For comparison, see \cite{doi:10.1021/nl5003526}.
}
% \caption{ \label{fig5}
%Instantaneous distributions of the real-valued electric field $\vE(\br,t)$ carried by the wave \eqref{Bessel}. At $y\r_0 %\approx 1$ the electric field rotates in the $yz$-plane.}
%
\end{figure}

\begin{figure}[ht!]
\centerline{\includegraphics[scale=3,clip=false,width=.7\columnwidth,trim = 0 0 0 0]{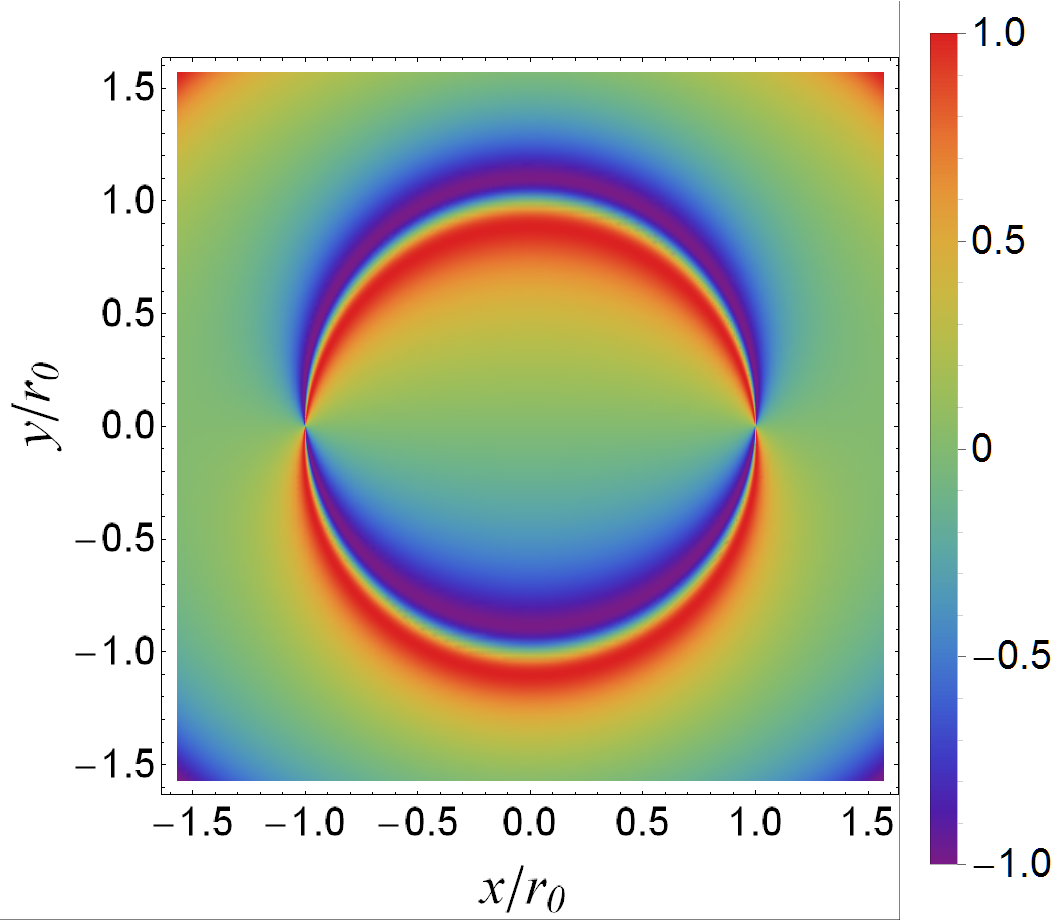}}
 \caption{ \label{fig1}
Ratio $S_3/S_0$  as a function of the scaled coordinates  $x/r_0$ and $y/r_0$. The electric field vector manifests a (local) transverse circular polarization at the points where $S_3/S_0=\pm 1$.
}
\end{figure}

\begin{figure}[hb!]
\centerline{\includegraphics[scale=3,clip=false,width=.7\columnwidth,trim = 0 0 0 0]{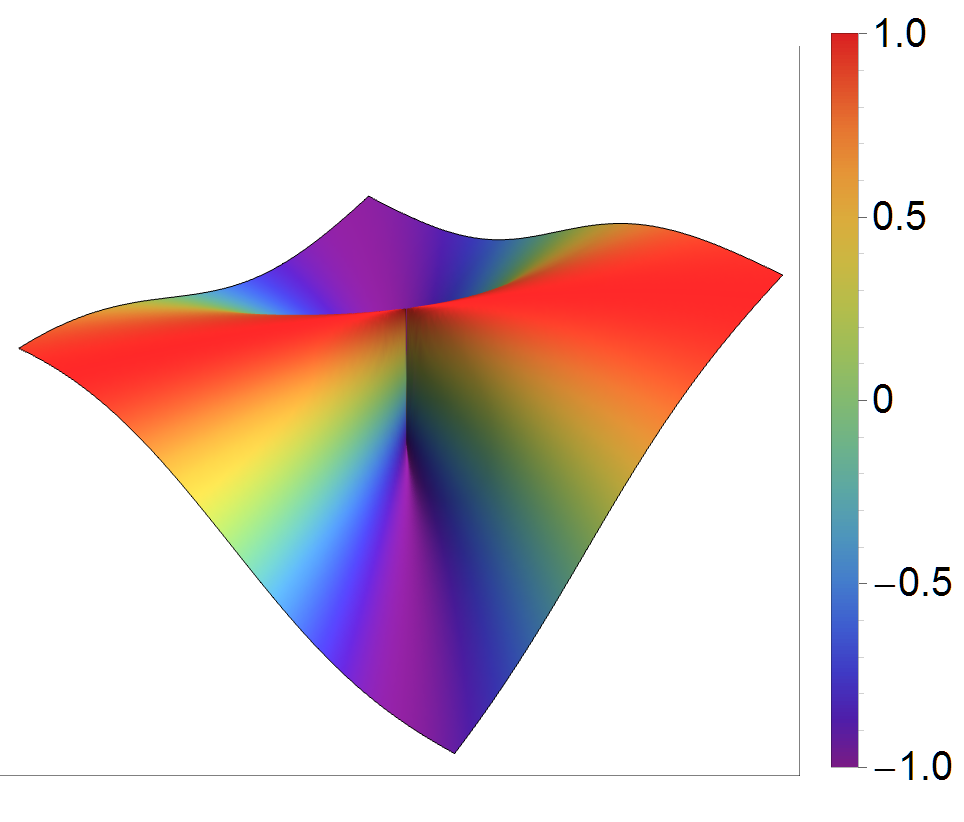}}
 \caption{ \label{fig1b}
Expanded 3D view of the  discontinuity  occurring  in  $S_3/S_0$ around  $x/r_0=1$ and $y \approx 0$.
}
\end{figure}

The polarization distribution on the $yz$-plane is characterized by the Stokes parameters
\begin{align}
S_0  =  & \; \alpha^2 + \beta^2, \label{S0} \\
S_1  =  & \; \alpha^2 - \beta^2 , \label{S1}  \\
S_2  =  & \; 0,\label{S2} \\
S_3  =  & \; 2 \alpha \beta, \label{S3}
\end{align}
where $\alpha \equiv \left({y}/{r}\right) k_r J_1(r k_r)$ and  $\beta \equiv  k_z J_0(r k_r)$. Equation \eqref{S2} shows that diagonal polarization never occurs in this kind of beams. The occurrence of transverse circular polarization is naturally marked by the conditions $S_1=0$ and $S_3 = S_0$. These are clearly satisfied when $\alpha = \beta$, namely when Eq. \eqref{CRBessel} becomes an identity. The  ratio $S_3/S_0$  is plotted as a function of the scaled coordinates  $x/r_0$ and $y/r_0$ in Fig.  \ref{fig1}. This figure shows that one has $S_3/S_0 = \pm 1$  in the tail of the  central core spot of the beam, namely for $x^2 + y^2 \approx r_0^2$.  Moreover, a  discontinuity  occurs around $x/r_0=\pm 1$ and $y \to 0$, as illustrated in detail in Fig. \ref{fig1b}.

\begin{figure}[ht!]
\centerline{\includegraphics[scale=3,clip=false,width=.7\columnwidth,trim = 0 0 0 0]{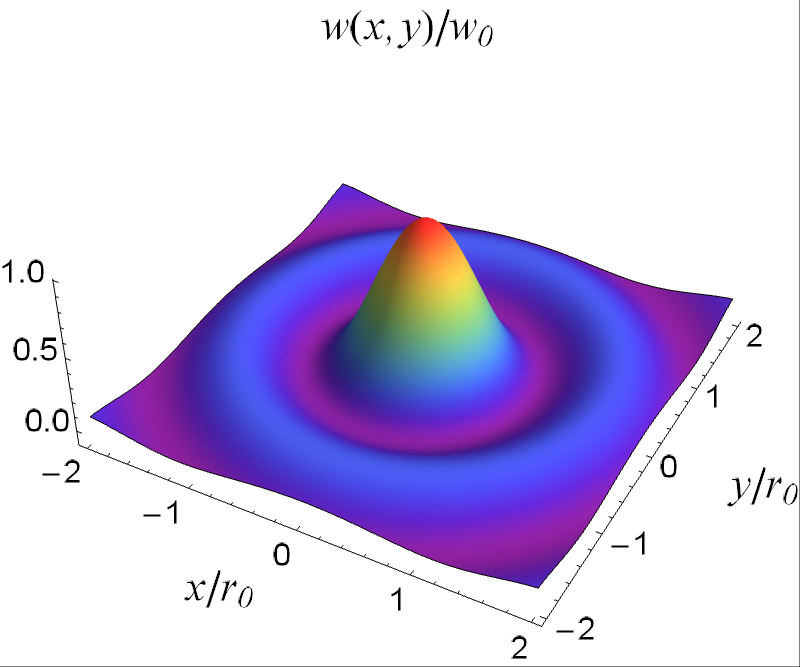}}
 \caption{ \label{fig2}
Time-averaged energy density $w(x,y)/w_0$ as given by Eq. \eqref{Energy}, with $r_0 = a/k_r$ and $\vartheta_0 = 15^\circ$.
}
\end{figure}

Despite of the unusual spatial distribution of  $S_3/S_0$, the projection of the beam upon the $xy$-plane displays the  typical Bessel form. This can be seen by calculating the  time-averaged energy density $w(x,y)$, which  is conserved along $z$ and is given by
\begin{align}\label{Energy}
w(x,y) = \frac{\epsilon_0}{4} \left( \vE^* \cdot \vE + c^2 \vB^* \cdot \vB \right),
\end{align}
where $\epsilon_0$ is the vacuum permittivity \cite{Jackson}. The plot of $w(x,y)$ normalized to the central value $w_0 \equiv w(0,0)$ is shown in Fig. \ref{fig2}.

\begin{figure}[h!]
\centerline{\includegraphics[scale=3,clip=false,width=1\columnwidth,trim = 0 0 0 0]{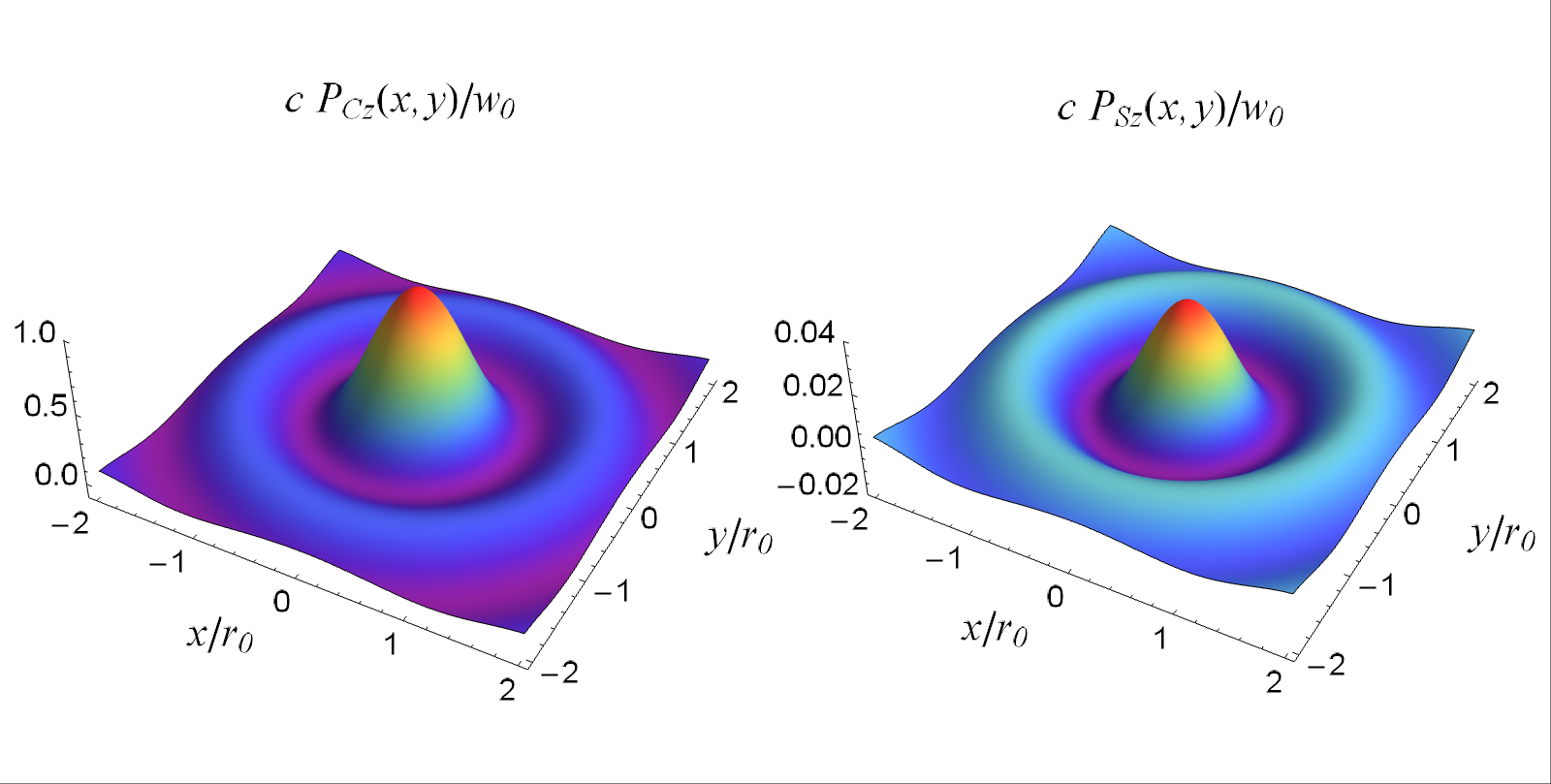}}
 \caption{ \label{fig3}
Distributions of the $z$-component of the time-averaged  canonical (left) and  spin (right) \emph{linear} momenta,  $\vP_C $  and  $\vP_S $ , respectively, as given by Eqs. (\ref{PC}-\ref{PS}). The occurrence of negative values for  $P_{Sz} $, should be noticed.
}
\end{figure}

Another relevant quantity that permits to characterize the beam, is the distribution of the linear momentum $\vP$ of the field, which naturally splits into a canonical and a spin part as   $\vP= \vP_C + \vP_S$, where
\begin{align}
\vP_C  =  & \; \frac{\epsilon_0}{4 \omega} \operatorname{Im} \left[ \vE^* \cdot \left( \bnabla \right) \vE + c^2 \vB^* \cdot \left( \bnabla \right) \vB  \right], \label{PC} \\
\vP_S  =  & \; \frac{1}{2} \bnabla \times \vS, \label{PS}
\end{align}
with
\begin{align}\label{SpinDen}
\vS  = & \;  \frac{\epsilon_0}{4 \omega} \operatorname{Im} \left( \vE^* \times \vE + c^2 \vB^* \times \vB  \right) \nonumber \\
\equiv & \; \vS_E + \vS_B,
\end{align}
being the spin AM density \cite{1367-2630-16-9-093037}. A straightforward calculation shows that $\vP_C = \hat{\bz} \, P_{Cz}$, $\vP_S = \hat{\bz} \,P_{Sz}$ and $\vS = \hat{\bx} \, S_x + \hat{\by} S_y$. This implies that $\vS \cdot \vP =0$, namely the helicity of the beam is identically zero:
\begin{align}\label{Helicity}
h  =   -\frac{\epsilon_0}{2 \omega} \operatorname{Im} \left( \vE^* \cdot \vB  \right) =0.
\end{align}
This peculiar phenomenon also occurs for surface-polariton fields and focused beams \cite{PhysRevA.85.061801,doi:10.1021/nl5003526}. The spatial patterns of $\vP_C $  and $\vP_S $ are shown in  Fig. \ref{fig3}. It is of worth noticing that $P_{Sz}$ takes negative values in some annular regions of the $xy$-plane, but the sum $P_{Cz} + P_{Sz} \geq 0$ is nonnegative everywhere. This peculiar phenomenon is well known for the transverse spin AM density in evanescent waves (see  \cite{PhysRevA.85.061801} and discussion later).

\begin{figure}[h!]
\centerline{\includegraphics[scale=3,clip=false,width=1.1\columnwidth,trim = 0 0 0 0]{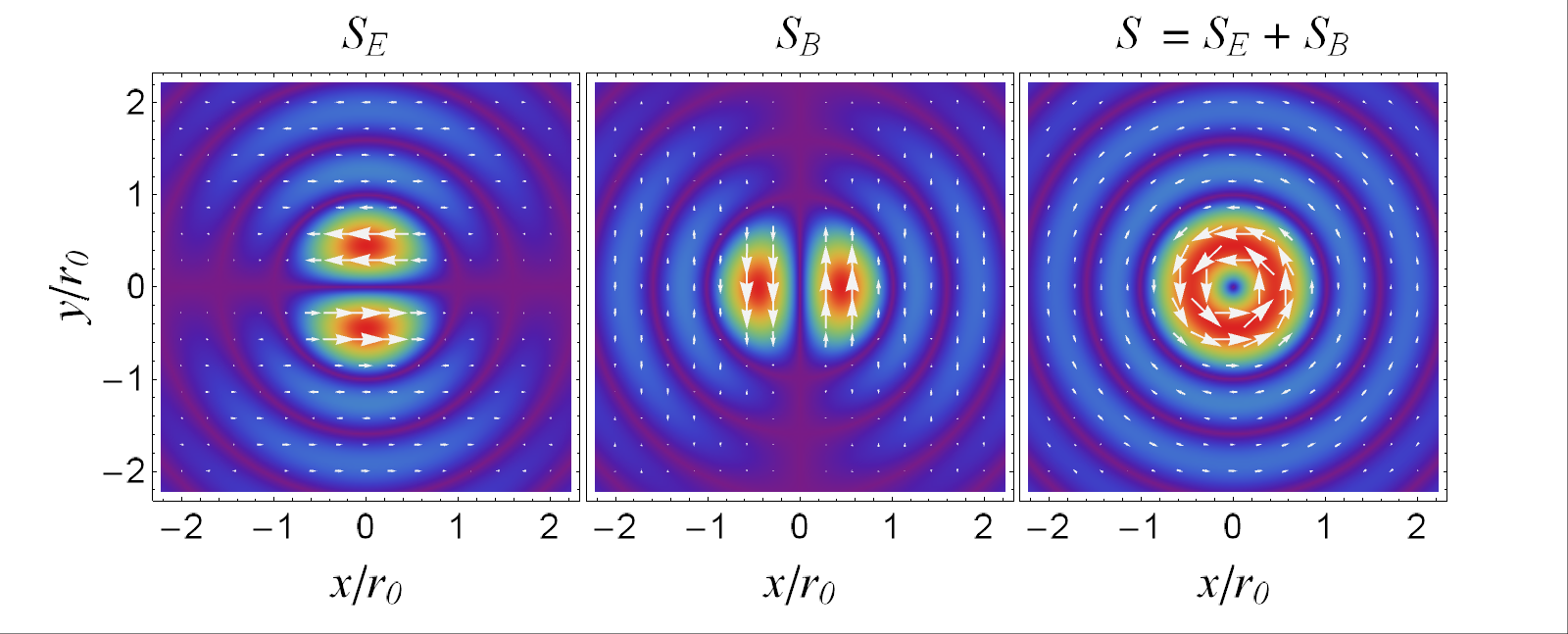}}
 \caption{ \label{fig6}
(Left to right) Distributions of the the spin AM densities vector fields $\vS_E$, $\vS_B$ and $\vS = \vS_E + \vS_B$ on the $xy$-plane, calculated from Eq. \eqref{SpinDen}. All plotted fields are normalized to $w_0/\omega$.
}
\end{figure}

Although the helicity \eqref{Helicity} of the wave is identically zero, the spin AM density is not. Since $\vS = \hat{\bx} \, S_x + \hat{\by} S_y$, it is clear that the spin AM for the wave \eqref{ERealBessel} is purely transverse. A straightforward calculation shows that $\vS_E = \hat{\bx} \, S_{Ex}$ and $\vS_B = \hat{\bx} \, S_{Bx} + \hat{\by} S_{By}$, with $\abs{S_{Bx}} \ll \abs{S_{By}} \approx \abs{S_{Ex}}$. Figure \ref{fig6} shows the vector field distribution on the $xy$-plane of (from left to right) $\vS_E$, $\vS_B$ and $\vS = \vS_E + \vS_B$, respectively, superimposed on a density plot of the  norm of the corresponding fields.
Both the electric and magnetic fields contribute to the transverse spin AM density, the two terms being in quadrature. As we shall show soon, the situation is drastically different from the one occurring with evanescent waves, where only the electric field contribution is nonzero.

\subsection{Evanescent waves}\label{Evanescentfield}

\begin{figure}[h!]
\centerline{\includegraphics[scale=3,clip=false,width=1.1\columnwidth,trim = 0 0 0 0]{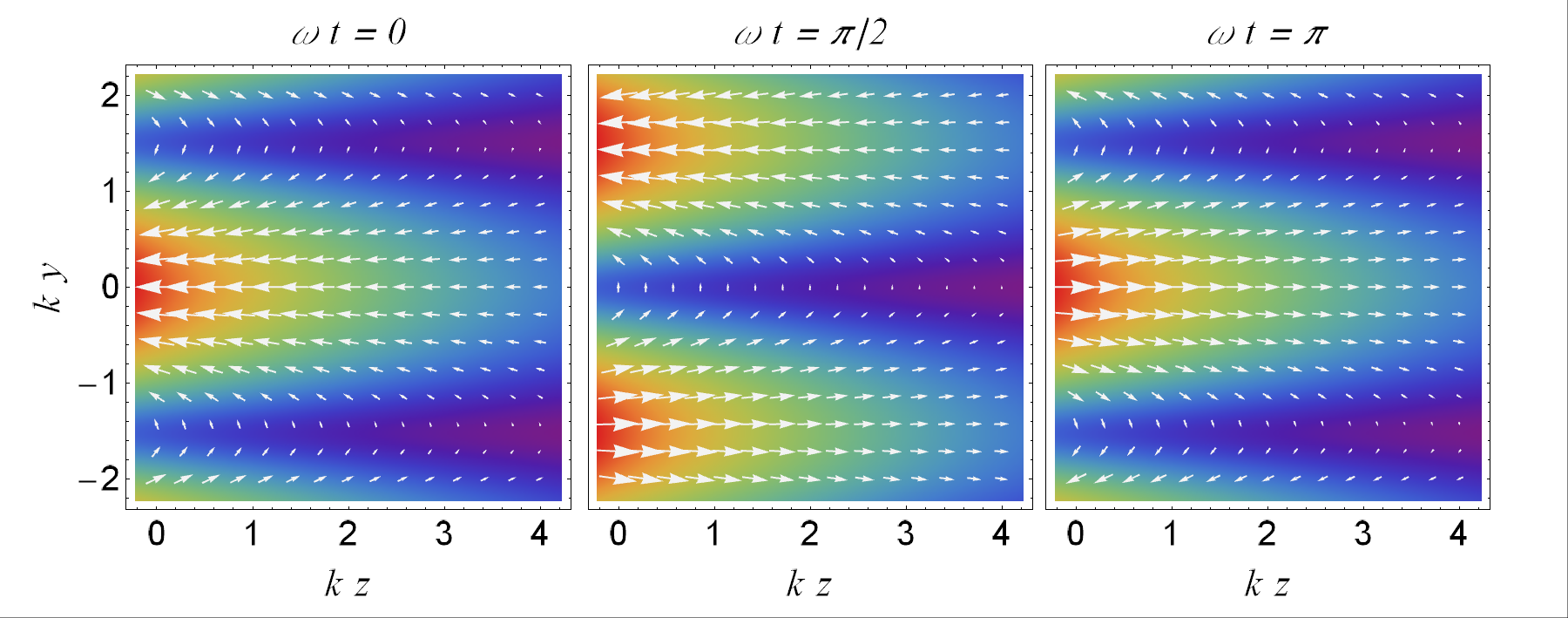}}
 \caption{ \label{fig7}
Instantaneous distributions of the real-valued electric field $\vE^R(y,z,t)$ as given in Eq.  \eqref{ERealEva} and evaluated for $\zeta = \pi/12$. In each point the electric field vector rotates clockwise.
}
% \caption{ \label{fig5}
%Instantaneous distributions of the real-valued electric field $\vE(\br,t)$ carried by the wave \eqref{Bessel}. At $y\r_0 %\approx 1$ the electric field rotates in the $yz$-plane.}
%
\end{figure}

Consider now the evanescent field  of an inhomogeneous plane wave exponentially decaying in the positive $z$ direction and propagating along the $y$ axis \cite{Jackson}:
\begin{align}\label{Evanescent}
\phi(\br,t) = \exp\left(- z k \sinh \zeta +i y  k \cosh \zeta  \right) \frac{e^{  - i \omega t }}{\omega},
\end{align}
where the real parameter $\zeta$ fixes the scale of inhomogeneity. For $\zeta = 0$ the wave is purely propagating. The electric and magnetic fields obtained by substituting $\phi(\br,t)$ into Eqs. (\ref{E}-\ref{B}), are given by
\begin{align}
\vE =  & \;  -c k^2 \phi \, \left( -i\hat{\by} \sinh \zeta + \hat{\bz} \cosh \zeta   \right), \label{EEva}\\
\vB =  & \;  - k^2 \phi \,  \hat{\bx}  \, \label{BEva}.
\end{align}
The corresponding real electric field is written as
\begin{align}\label{ERealEva}
\vE^R(y,z,t) =  & \; - k \exp \left( - z k \sinh \zeta \right) \nonumber \\
& \; \times \Bigl[ \hat{\by} \,\sin \left( y k \cosh \zeta - \omega t \right) \sinh \zeta \Bigr. \nonumber \\
& \; \phantom{ \times \Bigl[ i} \Bigl. + \hat{\bz} \, \cos \left( y k \cosh \zeta - \omega t \right)\cosh \zeta \Bigr] .
\end{align}
Figure \ref{fig7} shows the temporal evolution of $\vE^R(y,z,t)$ as a function of the scaled coordinates $k y$ and $k z$, superimposed on a density plot of the  norm of the field.

For an evanescent wave, the energy density depends on $z$ solely and for $z \geq 0$ it  decays as $w(z)/w_0 = \exp \left( - 2 z k \sinh \zeta \right)$. After a straightforward calculation, from Eq. \eqref{SpinDen} we obtain $\vS_B = 0$ and  $\vS_E = \vS$, where
\begin{align}\label{SAMEva}
\frac{\omega}{w_0} \, \vS = \hat{\bx} \exp \left( - 2 z k \sinh \zeta \right) \tanh \zeta.
\end{align}
This expression (apart from a different choice of the transverse axis) is in agreement with \cite{PhysRevA.85.061801}. The canonical and the spin momenta can be calculated by using Eqs. (\ref{PC}-\ref{PS}) and the result is
\begin{align}\label{SAMEva}
\frac{c}{w_0}\vP_C = & \;  \hat{\by} \exp \left( - 2 z k \sinh \zeta \right) \cosh \zeta, \\
\frac{c}{w_0}\vP_S  = & \; - \hat{\by} \exp \left( - 2 z k \sinh \zeta \right) \sinh \zeta \tanh \zeta.
\end{align}
Also these results are in agreement with Eqs. (8-9) in \cite{PhysRevA.85.061801}. It should be noted that the ``backward'' momentum  $\vP_S $ is compensated by the forward momentum $\vP_C $ and their sum is eventually forward:
\begin{align}\label{SAMEvaSum}
\frac{c}{w_0} \left( \vP_C+\vP_S \right) =   \hat{\by} \exp \left( - 2 z k \sinh \zeta \right) \operatorname{sech} \zeta.
\end{align}
Finally, the Stokes parameters are written as
\begin{align}
S_0  =  & \; k^2 \cosh( 2 \zeta) \exp \left( - 2 z k \sinh \zeta \right), \label{S0ev} \\
S_1  =  & \; k^2  \exp \left( - 2 z k \sinh \zeta \right) , \label{S1ev}  \\
S_2  =  & \; 0,\label{S2ev} \\
S_3  =  & \; - k^2 \sinh( 2 \zeta) \exp \left( - 2 z k \sinh \zeta \right), \label{S3ev}
\end{align}
with $S_3/S_0 = - \tanh(2 \zeta)$. This means that for $\zeta \gtrsim 1$ the electric field has (almost) transverse circular polarization  uniformly over the $yz$-plane.

\begin{figure}[h!]
\centerline{\includegraphics[scale=3,clip=false,width=1.1\columnwidth,trim = 0 0 0 0]{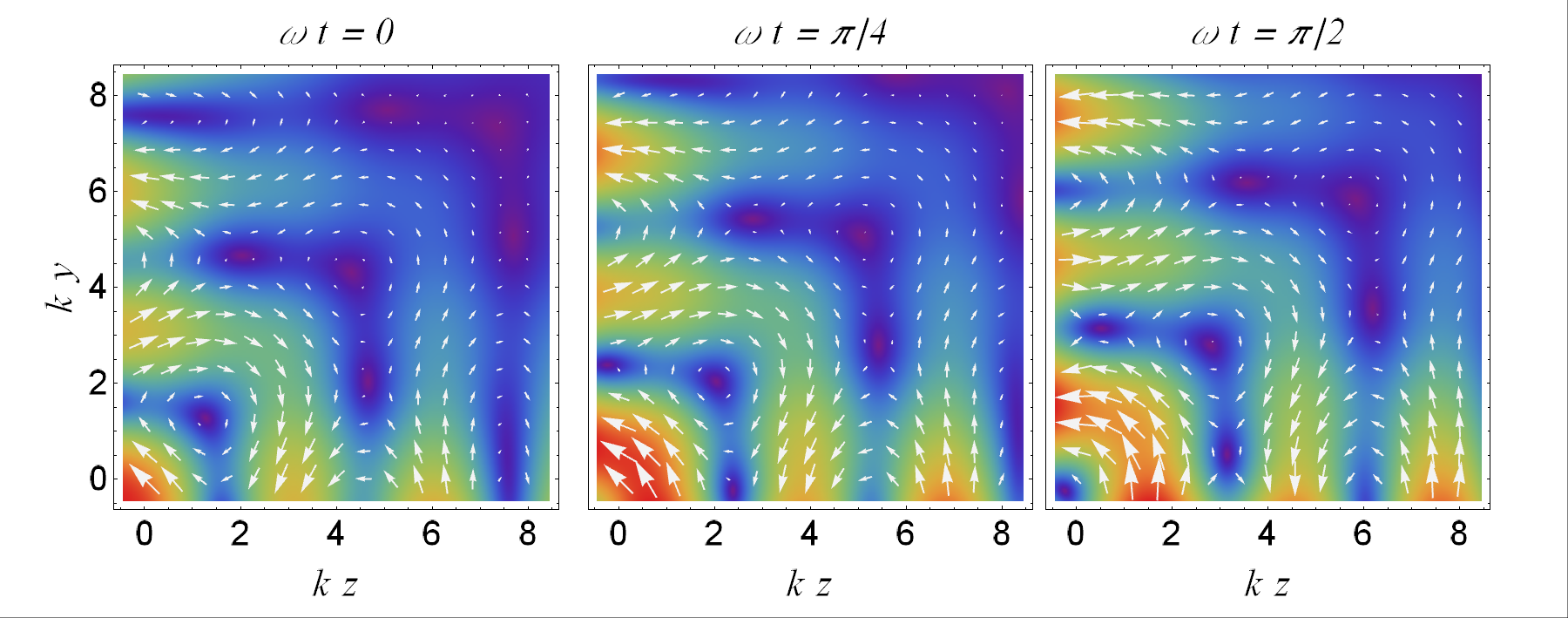}}
 \caption{ \label{fig8}
Instantaneous distributions of the real-valued electric field $\vE^R(y,z,t)$ generated by $\phi(\br,t) + \phi_\perp(\br,t)$ with  $\zeta = \pi/12$.
}
\end{figure}

More complex evanescent waves may be generated by interference. For example, suppose to illuminate the vertical edge of a parallelepiped immersed in a liquid with higher refractive index. Under suitable conditions, one  obtains inside the solid  two perpendicular evanescent waves, namely $\phi(\br,t)$  given in Eq. \eqref{Evanescent} and $\phi_\perp(\br,t)  = \exp\left(- y k \sinh \zeta +i z  k \cosh \zeta  \right) \exp( - i \omega t )/\omega$.
The time-evolution of the interference pattern generated by summing $\phi$ and $\phi_\perp$ is shown in Fig. \ref{fig8}.

\section*{Discussion and conclusions}\label{Conclusions}

The reason why transverse circular polarization is most easily observed in strongly focused beams, is evident from Eq. \eqref{E}. For a paraxial fundamental Gaussian beam with angular aperture {${\theta_0 \ll 1}$}, one has that $\partial_z {\psi} \sim 1 + O(\theta_0^2)$ and $\partial_y {\psi} \sim O(\theta_0)$. In this case the transverse and longitudinal components of the electric field vector do not have a comparable magnitude. Alternatively, one may consider using waveguides, where TM modes with transverse elliptical polarization are created by superposition of plane waves \cite{MIT,White}. As a matter of fact, it is sufficient to consider the interference of \emph{two} plane waves only, to obtain non-paraxial optical fields with nontrivial polarization patterns comprising transverse circular polarization, as noticed in \cite{Nesci} and \cite{Bliokhtwowaves}. 

For the Bessel field \eqref{Bessel} we have found that the orbital linear momentum is purely longitudinal, namely $\vP_C = \hat{\bz} P_{Cz}$. This implies that the orbital part of the angular momentum density is purely radial: $\mathbf{J}_O = \br \times \vP_C =(y \hat{\bx}  - x\hat{\by})P_{Cz}$ leading to a null total orbital angular momentum. Of course, one may think to carry out an analysis similar to the one presented in this work, but studying instead transverse orbital AM. We are currently investigating along this direction.

In conclusion, we have presented a perfectly general theory of light carrying an electric vector field circularly polarized in a plane containing the main axis of propagation. The novelty of our approach resides in its ``universal'' character, which provides for a unifying view of seemingly different wave propagation phenomena.
The success of such unification is made manifest in the two examples reported in this paper, where it is shown that for both propagating (Bessel) and evanescent (plane wave) fields, the underlying mechanism generating transverse spin AM is the same. Last but not least, our treatment reveals a somewhat hidden connection between light with circular polarization and the theory of complex functions.

\section*{Acknowledgment}\label{Acknowledgment}

We thank Konstantin Bliokh for  many fruitful discussions.

%
%\bibliography{biblio}

\begin{thebibliography}{22}
\bibitem [{Not()}]{NoteHelicity}%
  \BibitemOpen
  \href@noop {} {}\bibinfo {note} {Rigorously speaking, the helicity is defined
  as the projection of the spin angular momentum upon the direction of the
  linear momentum}\BibitemShut {NoStop}%
\end{thebibliography}

%
\end{document}